# Experimental study of intrinsic multiple Andreev reflections effect in GdO(F)FeAs superconductor array junctions

T.E. Kuzmicheva[1,2(a)], S.A. Kuzmichev[2], M.G. Mikheev[2], Ya.G. Ponomarev[2], S.N. Tchesnokov[2], Yu.F. Eltsev[1], V.M. Pudalov[1], A.V. Sadakov[1,3], A.S. Usoltsev[1], E.P. Khlybov[3,4], and L.F. Kulikova[4]

[1] *P.N. Lebedev Physical Institute of the RAS – 119991 Moscow, Russia*
[2] *M.V. Lomonosov Moscow State University –119991 Moscow, Russia*
[3] *International Laboratory of High Magnetic Fields and Low Temperatures – 53-421 Wroclaw, Poland*
[4] *Institute for High Pressure Physics of the RAS – 142190 Troitsk, Russia*

**Abstract –** We report the first observation of the intrinsic multiple Andreev reflections effect (IMARE) in S-n-S-…-S-arrays (S = superconductor, n = normal metal) formed by "break-junction" technique in GdO(F)FeAs superconductor ($T_C = 48 \div 53$ K). We show that superconducting gap peculiarities at dI/dV-spectra sharpen dramatically in the arrays as compared with that in the single-contact spectra; this enables to improve significantly accuracy of the bulk superconducting parameters determination. Using IMARE, we determined the large and the small gap values $\Delta_L = 11.0 \pm 1.1$ meV and $\Delta_S = 2.6 \pm 0.4$ meV. The BCS-ratio $2\Delta_L/k_B T_C^{local} = 5.0 \div 5.9 > 3.52$ ($T_C^{local}$ is the contact area critical temperature) evidences for a strong electron-boson coupling. The results obtained agree well with our previous data by Andreev spectroscopy for single SnS-contacts.

Among all newly discovered iron-based pnictide superconductors [1], ReOFeAs (Re = rare earth metals) or "1111-system", shows the highest critical temperature, up to $T_C \approx 56$ K [2]. Like high temperature superconducting cuprates, oxypnictides have a quasi-two-dimensional crystal structure that contains superconducting FeAs planes alternating along the *c*-direction with insulating ReO layers. The stoichiometric 1111 compounds are antiferromagnetic metals with a spin density wave ground state, whereas electron or hole doping turns the system into superconductivity [3,4].

Band structure calculations [5,6] showed four bands crossing Fermi level in the normal state, forming two concentric cylinder-like hole Fermi surface sheets at the Γ point of the first Brillouin zone and two quasi-2D electron sheets at the M points. These sheets may be considered as two effective quasi-2D bands [7], where two distinct superconducting condensates arise at $T < T_C$ [8]. The total density of states at the Fermi level is formed mainly by Fe 3d-states [9,10], suggesting these electrons to play a key-role in superconducting pairing. Interest in studying these compounds is related with still unknown and debatable paring mechanism and symmetry of the order parameter.

Unambiguous determination of the superconducting gap, Δ, remains a challenging issue for 1111-systems. The existing experimental results diverse significantly, including the ones measured by point-contact Andreev reflection (PCAR) spectroscopy (for a review see [11], and Table 1 in [12]). As for GdO(F)FeAs, no other data on Δ values are available until now beyond that reported in our previous works [12,13]. Lack of experimental information on Δ values for Gd-1111 evidently calls for their independent determination by other techniques. Here we report the first observation of intrinsic multiple Andreev reflections effect (IMARE) in Gd-based oxypnictides with natural S-n-S-n-…-S Andreev arrays (S = superconductor, n = thin normal metal layer with a thickness less than the quasiparticles mean free path; the latter means elementary S-n junctions to form a Sharvin-type contacts [14]). Using the spectroscopy based on IMARE technique we determined accurately two superconducting gap values, $\Delta_L$, and $\Delta_S$, and the corresponding BCS-ratios $2\Delta_{L,S}/k_B T_C$. We present also a comparison of our new IMARE spectroscopy data with earlier results of the SnS-Andreev spectroscopy [12].

The two sets of polycrystalline samples were used: fluorine-doped $GdO_{1-x}F_xFeAs$ with concentrations $x = 0.09 \div 0.21$ and $T_C^{bulk} = 48 \div 53$ K (EL- and KHL-series),

---

[(a)]E-mail: kute@sci.lebedev.ru

and oxygen-deficient GdO$_{0.88}$FeAs with $T_C^{bulk}$ = 52 ± 2 K. The polycrystalline samples were prepared by high-pressure synthesis detailed in [13,15]. The chips of Gd and As (99.9%) were placed in an evacuated quartz ampoule and held at T = 1050 °C during 24 hours to produce GdAs powder. Then high-purity single-phase GdAs, Fe$_2$O$_3$, FeF$_3$ and Fe powders were mixed together in nominal stoichiometric ratio and pressed into pellets of 3×3×3 mm$^3$. Afterwards, the pellets were placed in boron nitride crucible and synthesized under the pressure 50 kb and the temperature 1350 C during 60 min. The resulting polycrystalline samples characterization by X-ray diffraction showed the presence of prevailing superconducting phase with an admixture of FeAs (about 19%) and Gd$_2$O$_3$ (about 19%); measurements of the resistance and AC-susceptibility temperature dependences showed rather sharp superconducting transitions width about (1.5 ÷ 4.5) K [12,13,15]. The subsequent local energy-dispersive spectroscopy analysis (JSM-7001FA) has revealed that the incidental phases are concentrated in grains of about 1 μm size scattered in the bulk majority phase. This fact opens a possibility to probe properties of the true majority phase using local techniques.

Microcontacts were formed in the bulk of the studied samples by a "break-junction" technique [16,17]. The samples of typical dimensions about 2×1×0.1 mm$^3$ were attached to a springy sample holder by four contact pads made of liquid (at room temperature) In-Ga alloy; this set-up enables 4-contact measurements. The sample holder with the sample was further cooled down to T = 4.2 K. Gentle mechanical bending of the sample holder at 4.2 K generates a microcrack in the bulk of the sample. To operate in a superconductor-normal metal-superconductor (SnS) regime, the microcrack was precisely tuned until we achieved the desired Andreev-mode characteristic. The estimated width of the constriction is about 0.1 μm, therefore the applied bias current of order of 1 mA flowing through the constriction exceeds the critical current and turns the constriction into the normal state; the studied junction is equivalent to the superconductor-constriction-superconductor (ScS; c = normal metal or insulator) contact. The layered material (such as Gd-1111) is most likely exfoliates along the *ab*-plane, therefore in our setup a bias current flows along the *c*-direction. Location of the microcrack deep in the bulk of the sample prevents both, cryogenic clefts degradation, and overheating of the contact area by the bias current. Importantly, the "break-junction" technique enables to readjust mechanically the contact, promoting observation of several tens of various SnS-contacts during one and the same experiment.

In case of layered superconductor, exfoliation of the sample generates "steps-and-terraces" and thus may form not only single ScS-junctions but also arrays of the S-c-S-c-…-S-type. The latter represents several consequently connected single ScS-contacts. Fine mechanical readjustment provided by our set-up allows us to deal with ScS-arrays with various number *N* of junctions, usually up to 20.

Kümmel *et al.* showed [18] that multiple Andreev reflections effect [19] in symmetric SnS-contact manifests itself as an excess current in the low-bias region at the current-voltage characteristic (CVC) and a subharmonic gap structure (SGS) at the dI/dV-spectrum. For temperatures T < T$_C$ and for the ballistic limit [14], the SGS represents series of dynamic conductance dips at certain bias voltages

$$|V_n| = \frac{2\Delta}{en}, \qquad (1)$$

where $\Delta$ is the superconducting gap, $n = 1, 2,…$ – subharmonic order. Relative amplitude of these dips decreases as *n* increases [18]. Note that higher order Andreev peculiarities are much less probable. In case of a two-gap superconductor, one should expect the two such SGSs, corresponding to the large $\Delta_L$ and the small $\Delta_S$ gaps [20]. Generally speaking, in case of $\Delta_L/\Delta_S \geq 3$ the large gap SGS minima should not be visible in the range of $\pm 2\Delta_S/e$, therefore, only peculiarities caused by $\Delta_S$ are usually observed at the small bias voltages.

A dynamic conductance shown in Figs. 1, 2, and an excess-current CVC (see Fig. 2) are typical for clean classical SnS-contacts [18]; for these reasons, the theory by Kümmel *et al.* is applicable to results presented below. The main advantage of Andreev spectroscopy with symmetric SnS-junctions is that superconducting gap value(s) can be determined directly, using eq. (1), from the dI/dV spectra measured up to T$_C$, with no fitting parameters [18], in contrast to the case of asymmetric Andreev NS-contact [21]. The presence of several Andreev conductance dips in the dI/dV spectra (various *n* in Eq.(1)) facilitates precise determination of superconducting gaps.

Intrinsic multiple Andreev reflections effect, observed first in [22], takes place in Andreev arrays in layered superconductors, and is similar to intrinsic Josephson effect in S-I-S-I-…-S contact [23,24] (I = insulator). Since Andreev array is a stack of single SnS-junctions, its CVC and dynamic conductance spectra scale by a factor of *N* (where *N* is the number of contacts in the array); the SGS dips should occur at bias voltages

$$|V_n| = \frac{N2\Delta}{en}. \qquad (2)$$

Hence, by measuring dI(V)/dV-characteristics of Andreev arrays for different *N* and by normalizing them to a single SnS-contact dynamic conductance, it is easy to calculate the corresponding numbers *N* and the resulting superconducting gap value(s).

For the above reasons, and in order to check self-consistency of the measured superconducting gap values, we used both methods: Andreev spectroscopy of single symmetrical SnS-contacts and IMARE spectroscopy of Andreev arrays. In general, surface defects contribution to the dynamic conductance for an *N*-junctions array should be *N* times less than that for a single junction. In contrast to dynamic conductance peculiarities related to the bulk properties of superconductor, positions of surface-related gap peculiarities (if any) do not scale with *N*. Thus, IMARE spectroscopy provides high precision determination of the true *bulk* gaps.

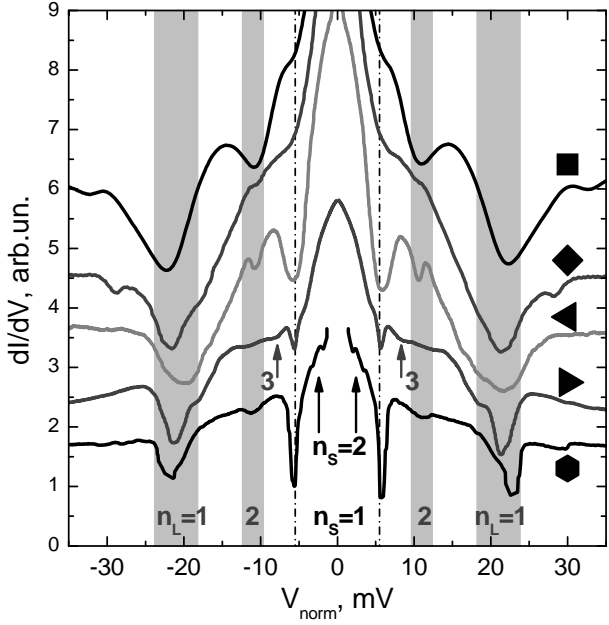

Fig. 1: Normalized dynamic conductance for single SnS-junction (■ EL1 sample, contact #d6, data from [12]), and for Andreev arrays of 2 junctions in the stack (♦ sample KHL3, contact #c), 6 junctions (◄ sample KHL9, contact #d07; ► oxygen-deficient sample KHL8, contact #d03), and 9 junctions (⬢ sample KHL12, contact #c). Curves are shifted vertically for clarity. The subharmonic gap structure for the large gap $\Delta_L \approx 11$ meV is marked by vertical stripes (the width reflects a 10% uncertainty) and $n_L$ labels; for the small gap $\Delta_S \approx 2.5$ meV – by dash-dotted lines and $n_S$ labels.

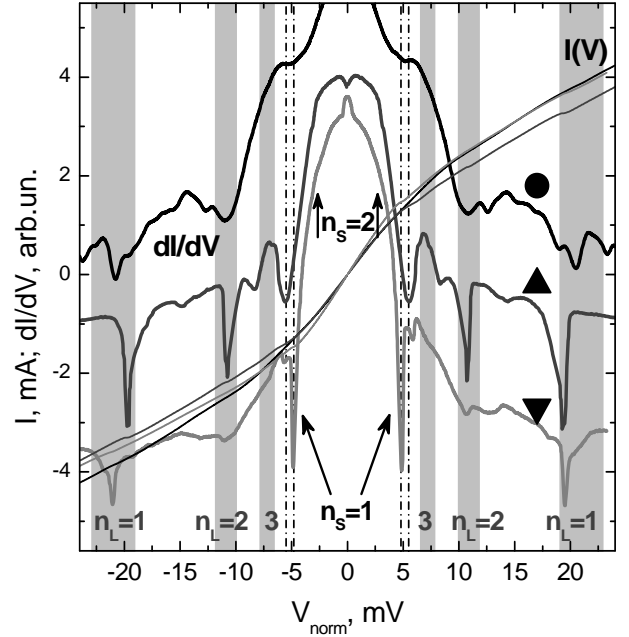

Fig. 2: Normalized to a single junction current-voltage characteristics and dynamic conductance for Andreev arrays of 9 junctions (● sample KHL7, contact #d02), 8 junctions (▲ sample KHL10, contact #e), and 11 junctions (▼ sample KHL9, contact #d). Curves are shifted vertically for clarity. Subharmonic gap structure corresponding to the large gap $\Delta_L \approx 10.9$ meV is marked by light gray vertical stripes (the width reflects a 10% uncertainty) and $n_L$ labels; for the small gap $\Delta_S \approx 2.6$ meV – by dash-dotted lines and $n_S$ labels.

Figure 1 shows a comparison between dynamic conductance for Andreev arrays obtained in this work and for single SnS-Andreev contact (upper black line, marked by square; data from [12]). At the latter spectrum, clear dips located at $V_1 \approx \pm 22$ mV and $V_2 \approx \pm 11$ mV, in accordance with eq. (1), can be considered as the first and the second Andreev minima for the large gap $\Delta_L \approx 11$ meV. The next peculiarity position around ± 5 mV is not related to the expected third harmonic position $V_3 \approx \pm 7$ mV for the large gap; for this reason we attribute it to the onset of the SGS defined by the small gap $\Delta_S \approx 2.5$ meV.

Other dynamic conductance curves presented in Fig. 1 correspond to 2-junctions array (KHL3 sample, contact #c, marked by rhomb), 6-junctions arrays (KHL9 sample, contact #d7, left triangle; oxygen-deficient KHL8 sample, contact #d3, right triangle), and 9-junctions array (KHL12 sample, contact #c, hexagon). Remarkably, the Andreev dips in the dI(V)/dV-characteristics for various arrays, coincide well with each other when the V-axis scaled down by factors of 2, 6, 6, and 9, correspondingly. One can see a good agreement between the large and the small gap values obtained by SnS-Andreev and IMARE spectroscopies. The coincidence of the values obtained by both methods leaves no doubt in the bulk gap nature of the observed SGS dips. Indeed, only for peculiarities caused by bulk effects the characteristic voltage may scale with number of junctions *N* in an array, thus proving the bulk origin of the SGS dips.

Figure 2 shows excess-current CVCs (thin lines) and dynamic conductance for Andreev arrays of 9 junctions (KHL7 sample, contact #d2, labeled by a circle), 8 junctions (KHL10 sample, contact #e, up triangle) and 11 junctions (KHL9 sample, contact #d, down triangle). After normalizing of all the curves to a single junction (and shifting them vertically for clarity), the minima positions nicely coincide. Gray vertical stripes depict bias voltages $V_1 \approx \pm 21$ mV, $V_2 \approx \pm 10.9$ mV and $V_3 \approx \pm 7$ mV, corresponding to the first, second and third Andreev peculiarities for the large gap; the stripe width reflects a 10% spread of the values. Dash-dotted lines and arrows depict peculiarities at $V_1 \approx \pm 5.2$ mV and $V_2 \approx \pm 2.6$ mV for the small gap (with 15% uncertainty). Using eqs. (1,2), we immediately obtain average gap values for three KHL samples: $\Delta_L \approx 10.9$ meV and $\Delta_S \approx 2.6$ meV. The details of spectra decoding for our stack SnS-contacts are presented further.

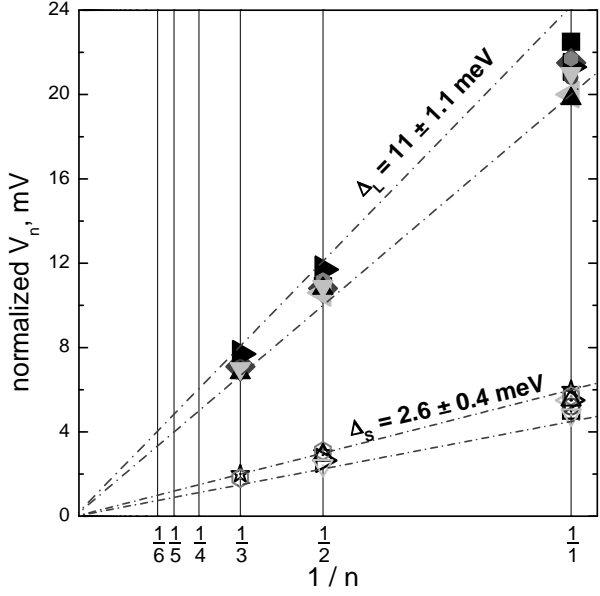

Fig. 3: Normalized to a single contact Andreev minima positions $V_n$ for both superconducting gaps versus inverse subharmonic order, $1/n$. Symbols correspond to those at the dynamic conductance curves in Figs. 1, 2. Star marks correspond to sample KHL8, contact #f (3 junctions in the stack) reported in [25]. Average values of the superconducting gaps are $\Delta_L = 11 \pm 1.1$ meV and $\Delta_S = 2.6 \pm 0.4$ meV. Dash-dotted lines show a 10% uncertainty for the large gap and 15% uncertainty for the small gap values.

Equations (1,2) imply a linear relation between the SGS minima positions $V_n$ and their inverse subharmonic number, $1/n$; the respective line should also tend to the (0; 0) point. The measured $V_n(1/n)$ dependence is presented in Fig. 3, where all $V_n$ values are normalized to a single junction. The gap peculiarity symbols in Fig. 3 repeat those for the conductance curves in Figs. 1, 2. The upper and lower dash-dotted lines indicate a 10% uncertainty interval for each gap. Clearly, the dip positions marked as $n_L = 1, 2, 3$ for the large gap follow the line, which passes through the (0; 0) point. Therefore, we believe, the peculiarities observed do satisfy eqs. (1, 2).

According to the theory by Kümmel *et al.* [18], the intensity of the SGS minima should decrease as subharmonic number *n* increases. The amplitude dependence could also help one to distinguish between the peculiarities belonging to the large and small gaps. In order to test this theory prediction, we scale the amplitude of various peculiarities to the amplitude of the $n_L = 1$ dip. The procedure of amplitude determination for Andreev minima is explained in Fig. 4. The characteristics plotted in the left pane of Fig. 4 are taken as typical data from Figs. 1, 2. Background dynamic conductance for the large gap SGS is shown by dotted lines in Fig. 4. These linear dependences are subtracted to exclude the background and thus to simplify the amplitude determination.

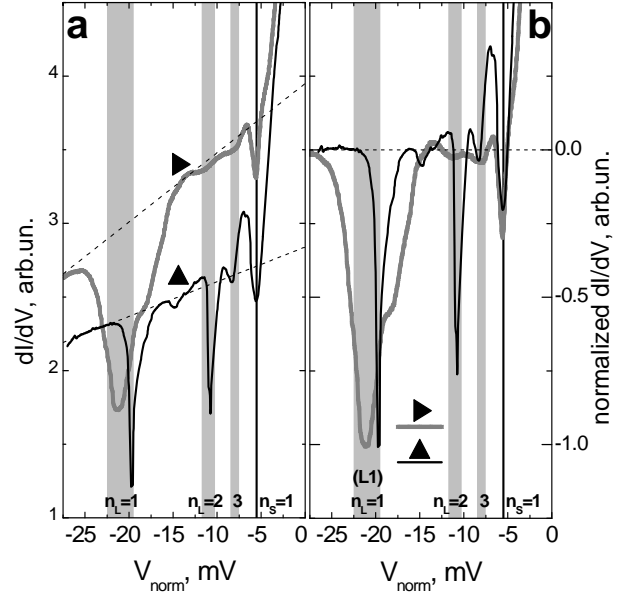

Fig. 4: **Left pane** (**a**) – dynamic conductance for SnS-Andreev contacts taken from Figs. 1, 2. Curves are shifted vertically for clarity. The subharmonic gap structure (SGS) for the large gap $\Delta_L \approx 11$ meV is marked by $n_L$ labels and gray vertical stripes (their width reflects 10% spread of data due to the difference in $T_C^{local}$ for the contacts); for the small gap $\Delta_S$ – by black vertical line and $n_S$ label. Dotted lines reflect a background dynamic conductance for the large gap SGS. **Right pane** (**b**) – dynamic conductance data from the left pane with subtracted background. The spectra are normalized to the amplitude of the first Andreev minima ($A_{L1}$) of the large gap SGS, taken as unity.

The dip amplitudes for small and large gaps were normalized then to the amplitude of the first Andreev minimum ($A_{L1}$) of the large gap SGS; the results are shown on the right pane of Fig. 4. In Fig. 5 we plotted relative amplitudes of the conductance peculiarities versus *n* for all spectra from Figs. 1, 2.

Figure 5 with a semi-log vertical scale clearly demonstrates nearly exponential decrease of the relative amplitude of the $\Delta_L$ dips ($n_L = 1, 2, 3$) with subharmonic number rise. As bias voltage decreases, a new sequence of peculiarities (marked as $n_S$) sets in with drastically enhanced $n_S = 1$ amplitude; in the same way, it decays to the next $n_S = 2$ minimum. Such non-monotonic amplitude behavior is the unambiguous evidence that the enhanced $n_S = 1$ dip amplitude is the onset of the small gap SGS. Another test of the association of the bias voltage minima with $\Delta_L$ and $\Delta_S$ harmonic numbers may be performed using eq. (2). The dip positions for the small gap, plotted in Fig. 3, also follow a line crossing the origin of coordinates; this supports grouping of the experimental points in Fig. 3 into two lines with different slopes. The two criteria described above prove that the peculiarities observed in dI/dV curves do form two distinct SGS corresponding to the large and the small gaps.

The two averaged gap values following from Fig. 3 are $\Delta_L = 11 \pm 1.1$ meV and $\Delta_S = 2.6 \pm 0.4$ meV. Single contact

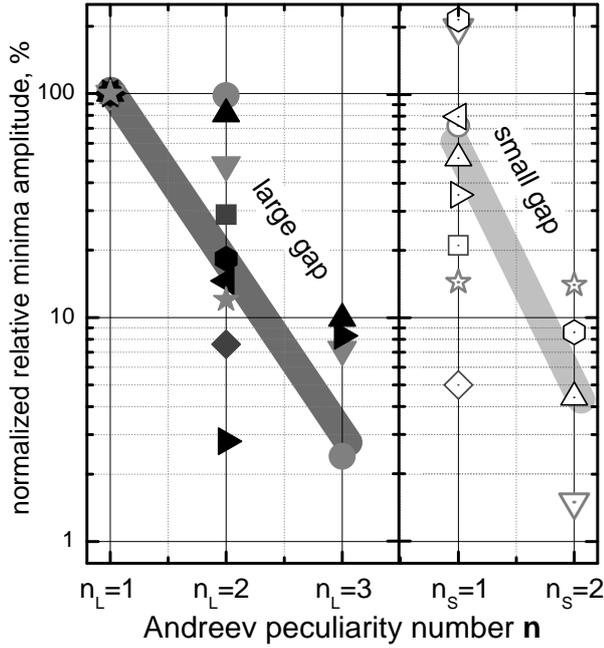

Fig. 5: Relative amplitudes $A_{L,S}$ of the SGS peculiarities attributed to the large ($n_L$ labels, solid symbols) and the small gap ($n_S$ labels, open symbols), normalized to $A_{L1}$ of the first Andreev minima for the large gap SGS (taken as 100%). Symbols correspond to those for the dynamic conductance curves in Figs. 1, 2. Star marks depict data for sample KHL8, spectrum for the contact #f (3 junctions in the stack) reported in [25]. Gray stripes are guides to the eye.

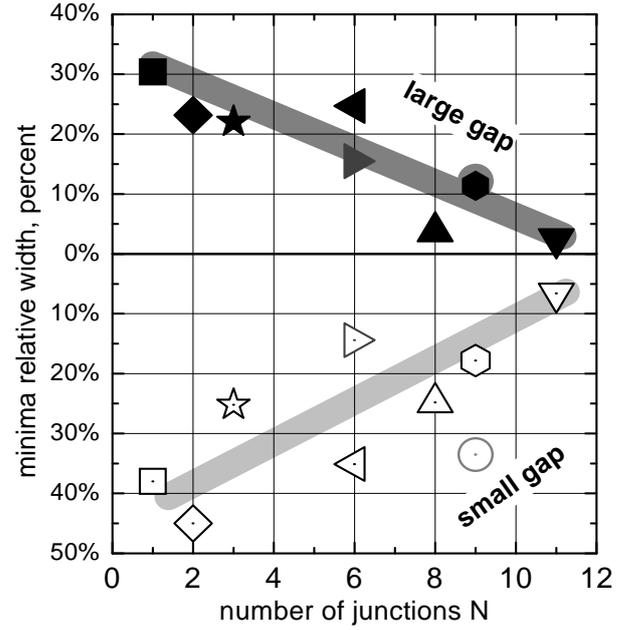

Fig. 6: The first Andreev minima ($n_{L,S} = 1$) relative half-width (normalized to a single contact, divided by its bias voltage $V_1$) versus number of junctions in the stack $N$ for the large gap (solid symbols, upper pane) and the small gap (open symbols, lower pane). Symbols correspond to those for the dynamic conductance curves in Figs. 1, 2. Star marks depict data for sample KHL8, spectrum for the contact #f (3 junctions in the stack) reported in [25]. Gray stripes are guides to the eye. Note, that vertical scales on two panels have opposite direction.

SnS-Andreev and IMARE spectroscopies also allow measuring a local critical temperature $T_C^{local}$. By "local" we mean the intrinsic critical temperature of the contact area (of the order of $0.1 \times 0.1$ μm$^2$) that can differ from the bulk $T_C$ obtained from resistance or AC-susceptibility measurements. For all studied contacts $T_C^{local}$ varies from 48 K to 50 K. $T_C^{bulk}$ value is usually greater than $T_C^{local}$, therefore the latter value is more appropriate for estimating the true BCS-ratio. Particularly, in our previous work [12] we used $T_C^{bulk}$ value and obtained, respectively, somewhat lower BCS-ratio $2\Delta_L/k_B T_C^{bulk} = 3.8 \div 5.8$. Here, using the $T_C^{local}$ value we find for the large gap $2\Delta_L/k_B T_C^{local} = 5.0 \div 5.9$, which exceeds the weak-coupling BCS-limit 3.52 and points at a strong electron-boson interaction in the "leading" (hole) bands with the large superconducting gap. This result agrees well with the BCS-ratio obtained by PCAR spectroscopy [26,27]. On the other hand, the BCS ratio for the small gap is much less, $2\Delta_S/k_B T_C^{local} = 1.0 \div 1.4$. Let us define the "intrinsic" critical temperature $T_C^S$ for the band with the small gap in a hypothetic case of zero interband coupling. Being smaller than 3.52, the obtained $2\Delta_S/k_B T_C^{local}$ ratio indicates that the superconductivity in the second ("driven") condensate at temperatures much higher than its $T_C^S$ is induced by the "leading" condensate. The driven band intrinsic $T_C^S$ may be evaluated by adopting the typical BCS ratio $2\Delta_S/k_B T_C^S = 3.5 \div 6$; as a result, we get estimate $T_C^S \approx (17 \div 10)$ K.

Noteworthy, by analysing shape of the Andreev dip, one can estimate type of the order parameter symmetry. As was shown theoretically in [28], the s-wave superconducting gap provides sharp, symmetrical SGS dips; on the contrary, in case of nodal gap the dips should be smeared and strongly asymmetric. Turning to the spectra presented in Figs. 1, 2, we attribute somewhat asymmetric minima $n_L$ to the extended s-wave rather than pure s- or d-wave symmetry type for the large gap. Similar conclusion can hardly be made for the small gap, because the corresponding $n_S$-peculiarities are observed on the strongly rising monotonic background.

In order to check whether the influence of surface defects is negligible in an array contact, we plotted in Fig. 6 relative width the of $n_{L,S} = 1$ SGS minima (for dI(V)/dV from Figs. 1, 2) versus corresponding number of junctions $N$ in the array. The relative width was calculated as its half-width divided by its bias voltage $V_1$. As expected, the relative half-width of the minima tends to decrease linearly as number of junctions $N$ increases. Seemingly, it happens because the contribution of surface-defects-influenced smearing of gap peculiarities falls with the increase in the number of contacts in the array $N$. Therefore, the IMARE spectroscopy regime enables to determine gap values more accurately.

As was mentioned above, in the "break-junction" technique, one obtains clean cryogenically cleaved surfaces located deep in the bulk of a sample. Nevertheless, Fig. 6 demonstrates that even in this case smearing of Andreev peculiarities is non negligible for the spectra of arrays with small *N*. The good agreement between our results for single contact SnS-Andreev reflections and IMARE spectroscopies (see Figs. 1, 2) evidences for high bulk homogeneity of the studied samples at the submicron scale.

In conclusion, we observed and studied intrinsic multiple Andreev reflections effect (IMARE) on natural S-n-S-n-…-S Andreev arrays made by the "break-junction" technique in GdO(F)FeAs polycrystalline samples with bulk critical temperatures $T_C^{bulk} = 48 \div 53$ K. The large $\Delta_L = 11 \pm 1.1$ meV and the small $\Delta_S = 2.6 \pm 0.4$ meV gap values were precisely measured by IMARE spectroscopy; the novel results obtained by the IMARE spectroscopy agree well with Andreev reflections spectroscopy data on single SnS-contacts published earlier [12]. The instrumental uncertainty of our set-up corresponds to about 0.1 meV dispersion of the gap values; the reported above 10-15 % uncertainty of the averaged $\Delta_{L,S}$ reflects the sample inhomogeneity. The measured gap values in combination with local critical temperatures correspond to the BCS-ratios $2\Delta_L/k_B T_C^{local} = 5.0 \div 5.9$, this value is consistent with the results of experimental works [26,27]; $2\Delta_S/k_B T_C^{local} = 1.0 \div 1.4$. These significantly different BCS ratios indicate that the large-gap condensate is in the strong-coupling regime and drives the superconductivity in the small-gap condensate for temperatures above $T_C^S \approx (10 \div 17)$ K. The shape of Andreev peculiarities in dI(V)/dV-spectra points to an extended s-wave type symmetry of the large order parameter. We also show that the array contacks with large number of junctions in the stack provide suppression of the surface defects influence (in comparison with single SnS-contacts); this suppression results in an essential increase of the accuracy of the gap measurements.

***

The research was supported by RFBR grant No. 12-02-31269, Russian Ministry for Education and Science (project No. 11.519.11.6012, grant No. 8375), and programs of the Russian Academy of Sciences.